# Electron energy-loss spectroscopy and ab initio electronic structure of the LaOFeP superconductor

Renchao Che, Ruijuan Xiao, Chongyun Liang, Huaixin Yang, Chao Ma, Honglong Shi, and Jianqi Li*,

*Beijing National Laboratory for Condensed Matter Physics, Institute of Physics, Chinese Academy of Sciences, Beijing 100080, China*

*Corresponding author. E-mail: <u>LJQ@aphy.iphy.ac.cn</u>

The electronic band structures of the LaOFeP superconductor have been calculated theoretically by the first principles method and measured experimentally by electron energy loss spectroscopy (EELS). The calculations indicate that the Fe atom in LaOFeP crystal shows a weak magnetic moment (0.14 $\mu_B$/atom) and does not form a long-range magnetic ordering. Band structure, Fermi surfaces and fluorine-doping effects are also analyzed based on the data of the density functional theory. The fine structures of the EELS data have been carefully examined in both the low loss energy region (< 60 eV) and the core losses region (O K, Fe $L_{2,3}$, and La $M_{4,5}$). A slight bump edge at ~44 eV shows notable orientation-dependence: it can be observed in the low loss EELS spectra with q//c, but becomes almost invisible in the q⊥c spectra. Annealing experiments indicate that low oxygen pressure favors the appearance of superconductivity in LaOFeP, this fact is also confirmed by the changes of Fe $L_{2,3}$ and O K excitation edges in the experimental EELS data.

**I. Introduction**

The superconductivity of LaOFeP was first discovered in 2006 by Kamihara *et al.* [1], although it had been synthesized early in 1995 by Zimmer *et al* [2]. Very recently, La[$O_xF_{1-x}$]FeAs was found to be a superconductor with $T_c \sim$ 26 K. In a structural point of view, the higher superconducting $T_c$ in La[$O_xF_{1-x}$]FeAs is achieved by replacing P with As and partial substitution of $O^{2-}$ by $F^-$ in LaOFeP [3-7]. One of the arresting characteristics of this type of materials is the occurrence of superconductivity at ~4K [1] with the presence of Fe, since ferromagnetism and superconductivity are considered to compete in conventional superconductors, although in principle any metal might become a superconductor in its non-magnetic state at a sufficiently low temperature [9]. Iron has 3*d* transition electrons and exhibits ferromagnetic character. Therefore, LaOFeP differs possibly from conventional superconductors. The number of iron-based superconductors identified so far is still limited. Three examples are iron under extreme pressures (15-30 GPa), filled skutterudite $LnFe_4P_{12}$ (Ln=Y, La) that has been prepared at high temperatures and high pressures [9-11] and ($La_{1-x}Sr_x$)OFeAs [26]. The discovery of superconductivity in LaOFeP provides a new case for researching a superconductor containing a ferromagnetic element. Actually, the strong electron correlation, coexistence and competition between superconductivity and ferromagnetism have been extensively discussed as a critical issue for many years.[12-14]

The understanding of the electronic structure is important to obtain insight into the superconductivity mechanism. Correlation between the electronic structure and superconductivity has attracted intensive interest. LaOFeP is an iron-containing superconductor and is prepared through a solid state reaction method or a high-temperature process together with an arc melting process [1,8]. S. Lebègue investigated the electronic structure of LaOFeP by means of *ab initio* calculations using density functional theory (DFT) [15]. His research focused on the charge transfer in this layered material and suggested that the La-O and Fe-P intra-layer bonding present a significant covalency, whereas the inter-layer bonding is ionic. A significant two-dimensional character of this system is found. In our previous work, the high quality LaOFeP superconducting materials were synthesized by a two-step solid reaction method [8]. In the present work, we report on the electronic band structure and magnetic moment of LaOFeP calculated by the first principles calculations. The effect of the atomic position of P on the magnetic moment is also discussed. We have observed certain EELS features from LaOFeP that can be attributed to the collective plasmon excitations, interband transition, or core-level transitions. Orientation-dependence for typical edges in LaOFeP are also examined in comparison with theoretical simulations. Moreover, annealing experiments at 1200ºC confirm that low oxygen pressure is a key factor for the occurrence of superconductivity in LaOFeP. Since the F-doped LaOFeP exhibits higher superconductivity $T_c$ [1-5], the influence of F-doping on electronic structure and Fermi surface is briefly discussed. We suppose our results might shed light on the understanding of the superconductivity mechanism of LaOFeP.

**II. Experiment and computation details**

The LaOFeP polycrystalline sample used in this study was synthesized by a method similar to the one reported in Ref. 8. The starting materials with appropriate composition were mixed, ground thoroughly and pressed into pellets. The pellets were enclosed inside an evacuated quartz tube. The vacuum of the tube was better than $1\times10^{-5}$ Pa. High purity Ar gas was filled to prevent implosion of the silica tubes. The first calcinations were done at 1200ºC for 12 h (sample without annealing). Then, the pellet was ground into powder and pressed into a pellet again for the second period of calcination (sample with annealing). High purity Ar gas and ~0.1 g $La_2O_3$ powder (to supply oxygen source) was added into the silica tube. The sintering temperature and time were the same as the first time and sample B was obtained. The purpose of the second sintering was to clarify the effect of low oxygen annealing on the superconductivity improvement of LaOFeP.

An FEI Tecnai F20 transmission electron microscope (TEM) equipped with a post-column Gatan imaging filter (GIF) was used for EELS measurement and TEM imaging. The energy resolution was ~0.85 eV, determined by the full-width at half-maximum (FWHM) of the zero-loss peak. To avoid electron-channeling effects, the selected grain was tilted slightly off the zone axis by 1-2º. The convergence angle was ~0.7 mrad (q≈0.04 Å$^{-1}$) and the collection angle was ~3 mrad (q≈0.17 Å$^{-1}$). The crystal structure of LaOFeP for the band structure calculation is based on the tetragonal structure with lattice parameters a=b=3.964 Å and c=8.512 Å and space group of P4/nmm [1-2]. The full potential linearized augmented-plane-wave+local orbital (APW+LO) were used as implemented in Wien 2k code [16]. The generalized gradient approximation (GGA) and the GGA+spin were used to treat the exchange and correlation effects in non-magnetic and magnetic states, respectively. The $R_{mt}K_{max}$ was set to be 7.0 to determine the basis size for spreading the plane wave. For the self-consistent calculations in all three cases, the energy, charge and force are converged into $10^{-5}$ Ry, $10^{-4}$ e and 1 mRy/au, respectively. The self-consistency was carried out on 4000 total k-points, which corresponded to a 20×20×9 k-mesh in the irreducible Brillouin zone (BZ). The Fermi surfaces (FS) for LaOFeP and F-doped LaOFeP were calculated on a 27×27×12 mesh containing 10,000 k-points in total and viewed by the XCrySDen package [17]. Theoretical data of core-loss EELS and low-energy spectra were simulated by the TELNES program and OPTIC program within the Wien 2k code. For the F-doped LaOFeP system, the F ion was considered to randomly occupy the O position, so the virtual crystal approximation was used to simulate the effect of F-doping. In the F-doped system, 5000 k-points were used in the self-consistency to obtain the convergence of total energy.

The electronic correlation is indeed important in transition metal oxides. In fact, we have considered the strong electronic correlations in this system using the LDA+U method, and we find that the result are hardly dependent of the magnitude of the electronic correlations due to its weak magnetic moment. In addition, our careful calculations using the LDA+U method show that the electronic correlation for Fe 3d electrons gives rise to only small modifications in the results and that LaOFeP is considered to be in weakly correlated regime, in well agreement with the LDA+DMFT (dynamics mean field theory, QMC) study as reported recently [28]. So, we carried out the calculations for the electronic structure as well as the electron energy loss near edge structure (ELNES) without considering the electronic correlation.

**III. Results and Discussion**

In order to systematically investigate the impact of the magnetic order and electronic structure in the LaOFeP system, three kinds of magnetic states were calculated, which includes the non-magnetic (NM), ferromagnetic (FM) and antiferromagnetic (AFM) state, as shown in Fig. 1. In the AFM state, we constructed an AFM ordering by flipping the spin orientation of one Fe atom in the cell, as illustrated in Fig. 1(c), in which the space group was changed into $P\bar{4}m2$. The heights of La and P were respectively relaxed in each model. The calculated results for the three kinds of magnetic states indicate that the ground state of LaOFeP is a weakly magnetic state. The magnetic moment of an Fe atom calculated in the FM state is 0.14 $\mu_B$/atom. The calculation of the AFM state almost converges into an NM state with a magnetic moment of 0.006 $\mu_B$ for each Fe atom. Although the calculated total energy of the three states is quite similar, the FM state has the lowest total energy of the three. Figure 2 shows the calculated density of states (DOS) near the FM level. It is can be seen that the main contribution arises from the Fe-P layer [15]. It should also be noted that the profile of spin up DOS and spin down DOS for Fe atom are almost symmetric, revealing that the Fe atom does not form a long-range magnetic ordering in this system.

Figure 3(a)-(c) shows the electronic band structural features of LaOFeP. Figure 3(a) is a brief energy level scheme for LaOFeP, qualitatively illustrating the bonding states and anti-bonding states from -20 eV below $E_F$ to 3.0 eV above $E_F$. Within deeper binding energy regions (-16 ~ -20 eV and -14 ~ -16 eV), the electronic states are mainly composed of LaO entity. Figure 3(b) and 3(c) are the spin-up and spin-down band structures for LaOFeP from -6.0 eV to 3.0 eV (Fermi energy at 0 eV).

Fundamental conductivity property is governed by the electronic states near the Fermi level. In the case of LaOFeP, hybridization states composed of Fe 3d with P 3p have large weights and represent the main contribution to the density of states (DOS) near Fermi energy. Referring back to the DOS in Fig. 2, we can identify that the 10 bands near the $E_F$ ranging from -2.5 eV to 2.5 eV are mainly composed of Fe 3d states. Detailed analysis of the atomic orbital decomposition from wave functions is performed according to the crystal field theory. One Fe atom and four P atoms form a FeP$_4$ tetrahedron. Two P atoms are located upward from the Fe atom layer and make a P-Fe-P bonding angle of 120.18º. The other two P atoms are located downward from the Fe atom layer and form a P-Fe-P bonding angle of 104.39º. Hence, this distorted FeP$_4$ tetrahedron is suppressed along the z axis of LaOFeP. The Fe 3d bands are crystal field split in the tetrahedral P environment. Triple degenerate $t_{2g}$ states ($d_{xy}$, $d_{xz}$, $d_{yz}$) are located above the doubly degenerate $e_g$ states ($d_{z2}$, $d_{x2-y2}$). If we set the c axis of LaOFeP as the z direction of the FeP$_4$ tetrahedron, disposal calculation of Fe 3d orbital shows that the $dz^2$ orbital extends along the c axis of LaOFeP and both the $d_{x2-y2}$ and $d_{xy}$ orbitals disperse within the a-b planes. Sub-bands of Fe 3d and P 3p intersect with the Fermi level. These partially filled bands might play an important role in the superconductivity and provide an opportunity for the improvement of superconductivity induced by hole-doping or electron-doping. It can be identified from Fig. 3 that there are two $e$ bonding orbitals ($d_z^2$ and $d_x^2{}_{-y}{}^2$) about -2.5 eV~-1.0 eV below the $E_F$ and two $e$ anti-bonding orbitals (also $d_z^2$ and $d_x^2{}_{-y}{}^2$) about 0~0.8 eV above $E_F$. Above these two orbitals, i.e., in the range from -1.0 eV to 0 eV and from 0.4 eV to 2.5 eV, the dominating part is the bonding and anti-bonding orbitals of three $t_2$ orbitals ($d_{xy}$, $d_{yz}$ and $d_{xz}$). Below the $E_F$, five Fe 3d bonding orbitals are completely occupied, one Fe $d_z^2$ anti-bonding orbital and two $t_{2g}$ anti-bonding orbital are partial occupied, so there are nearly six electrons filling in the Fe 3$d$ orbitals. The ionic charge state of Fe in this compound is nearly Fe$^{2+}$. The 3d electrons in two equivalent Fe atoms occupy these orbitals by 50% probabilities respectively, resulting in symmetric spin up DOS and spin down DOS, so the whole system shows very weak magnetic moment.

Our calculations also revealed that the height of the P atom inside LaOFeP lattice influences the magnetic moment of Fe atom notably. For example, when we use the experimental height of the P atom in the calculations, the length of the Fe-P bond is 2.286 Å and the calculated magnetic moment of Fe is 0.08 $\mu_B$/atom; when we use an optimal height of P atom in the calculations, the length of the Fe-P bond changes to 2.243 Å, a decrease of ~2% compared to the experimental value, but the magnetic moment of Fe increases to 0.14 $\mu_B$/atom. Further adjusting the height of the P atom, the magnetic moment of Fe might completely disappear. We also investigated the influence of the height of a La atom on the magnetic moment of this system, but no distinct effect was observed. This fact indicates that it is the specific structure of Fe-P tetrahedron that induces the low magnetic moment of Fe in this material. Y. Kamihara et al.[1] mentioned that the superconductivity in LaOFeP is drastically destroyed by varying Fe to Mn ($3d^5$) and Co ($3d^7$), in which the LaOMnP shows semiconductor characteristics while LaOCoP is a metal. We also calculated the electronic structure of LaOMnP and LaOCoP. They all show notable differences from LaOFeP. The most remarkable characteristic is the high magnetic moment of transition metal cations. The calculated magnetic moments of Mn and Co are 1.33 $\mu_B$/atom and 0.56 $\mu_B$/atom in LaOMnP and LaOCoP, respectively. In these materials, the magnetic scattering destroys the superconductivity, which implies that the very low magnetic moment of Fe atoms in LaOFeP is a necessary condition for the appearance of superconductivity in LaOFeP.

As shown in our calculations, the Fe ion possesses weak magnetic moment in contrast with the counterpart in other iron oxides, and the magnitude of the magnetic moment of iron ions is sensitive to the height of P atoms inside Fe-P tetrahedron. So, it is not surprising that the competing spin fluctuation would be controlled through changing the Fe-P bonding length by F doping or element substitution, and this antiferromagnetic spin fluctuation is evidenced by the newly discovered spin density wave instability which can be suppressed by F-doping in LaOFeAs compound[27].

Electron states around the Fermi surface affect the superconductivity measurably, especially in the case of hole-doping or cation-doping high-temperature cuprate superconductors (HTS).[18-22] We investigated the Fermi surfaces in LaOFeP to try to explore the origination of its superconductivity. According to the band structure of LaOFeP in Fig. 3(b) and 3(c), there are five energy levels across the $E_F$ in both spin up bands and spin down bands, which correspond to five Fermi surfaces in spin up state and spin down state respectively, as shown in Fig. 4 and Fig.5. Two cylindrical FSs centered along the A-M direction (as shown in Fig. 4(c) and Fig. 5(c)) exist in both spin up and spin down orientations. According to the band structure of LaOFeP in Fig. 3, these two Fermi surfaces are attributed to three $t_{2g}$ orbitals. That is to say, they mainly have the characteristics of Fe $d_{xy}$, $d_{yz}$ and $d_{xz}$. The other three FS are centered by the Z-Γ direction, in which the larger one is attributed to $d_{z2}$ orbital, while the two smaller FSs (as shown in Fig. 4(c) and Fig. 5(c)) have the characteristics of the $d_{xz}+d_{yz}$ orbital. This is different from the high-$T_c$ cuprates, in which the electrons near the $E_F$ are mostly occupied by Cu $d_{x2-y2}$ states. This kind of difference is mainly caused by the different coordination condition between Fe-P and Cu-O. The former is tetrahedral, while the later is octahedral. Comparing the Fermi surfaces in the spin up and spin down orientation, it should be noted that the weak magnetic moment of the Fe atom originates from the electrons in the $d_z^2$ orbital. Not only are the Fermi surfaces corresponding to $d_z^2$ orbital (as shown in Fig. 4(a) and Fig. 5(a)) obviously different in their spin up and spin down orientations, but also the position of their energy levels containing $d_z^2$ characteristics near the Z and Γ points are different – the spin down $d_z^2$ orbital is about 0.2 eV higher than the spin up $d_z^2$ orbital.

Recently, Y. Kamihara et. al[1] reported that the F-doped LaOFeP exhibits higher superconductivity $T_c$ (~5.5K) and F-doped La[$O_{1-x}F_x$]FeAs shows a $T_c$ value at 26 K[3], so we use the first principles calculations to investigate the influence of F-doping on the electronic structure and Fermi surface (FS) in the LaOFeP system. It is considered that the doped F atom occupies the position of the O atom. We take the method of virtual crystal approximation to study the electronic structure of La($O_{0.9}F_{0.1}$)FeP. To understand the effect of F-doping, the electronic structures of the doped and un-doped LaOFeP are compared. Figure 6 is the band structure of La($O_{0.9}F_{0.1}$)FeP. Since F-doping mainly influences the states near the Fermi level, only the bands ranging from -1.0 eV to 1.0 eV are considered. For convenient comparison, the band structure of LaOFeP is shown by dashed lines. F-doping induced more electron-doping in the system, so the Fermi level shifts to the higher energy direction. According to Fig. 6, the introduced electrons are mainly occupied into the $d_z^2$ orbitals near the Z and Γ point, which makes the energy level near the Z and Γ points decrease ~0.1 eV compared with the un-doped material. This kind of alternation also can be observed in the Fermi surface (FS) of La($O_{0.9}F_{0.1}$)FeP as shown in Fig. 7 and Fig. 8. Comparing these with the FS of LaOFeP in Fig. 4 and Fig. 5, we can see that the two cylindrical Fermi surfaces centered on the A-M direction changed little, but the FS centered along the Z-Γ direction, especially the spin down $d_z^2$ characteristic FS (Fig. 8(a)) changed greatly. For spin up orientation, the larger FS centered on the Z-Γ direction changes into a small one, and the smallest FS completely disappears when F is doped; for the spin down orientation, the cylinder radius of these FS (Fig. 8(a)-(c)) is shorter than for the un-doped LaOFeP. This is because the electrons introduced by F atoms occupy the $d_z^2$ orbital of Fe atom, so the unoccupied energy levels near Z point shift to the low energy direction and change into occupied levels. For the spin up orientation, the shifted levels do not cross the Fermi level in the Z-R direction anymore, so the corresponding FS disappears; for the spin down orientation, the position of these levels decreases, making the cross point between these level and $E_F$ shifts toward the Z point, so the cylinder radius of corresponding FS reduces. The calculations of magnetic moment for the doped and un-doped system indicate that, when 10% F atoms are doped into LaOFeP system, the magnetic moment of a cell changed to 0.24 $\mu_B$/cell, and the magnetic moment of an Fe atom is 0.1 $\mu_B$/atom, which is less than in the un-doped material. The difference in electronic structures between La($O_{0.9}F_{0.1}$)FeP and LaOFeP indicates that F-doping introduces electrons into the Fe $d_z^2$ orbital so as to adjust the carrier concentration in the FeP layer.

Next, the anisotropic properties of electronic states in LaOFeP will be emphasized. EELS analysis is performed and the correlation with the electronic band structure of LaOFeP is established. Figure 9(a) shows the crystal structure model of LaOFeP. Two-dimensional feature of LaOFeP can also be confirmed from the SEM observation and high-resolution TEM imaging, as shown in Fig. 9(c) and (d). Figure 9(b) shows the electron energy loss spectra in the low energy range recorded from the LaOFeP superconductor with the incident electron beam parallel with and perpendicular to the c-axis, respectively. The main difference between the two spectra is peak c at around 43.5 eV. Peak c shows a slight bump feature. It is expected that the spectrum measured with momentum transfer q//c should reveal more interlayer bonding contribution than the spectra recorded with q⊥c.

In the q⊥c spectra, the data should show a major contribution from the intralayer bonding feature such as, La-O bonding and Fe-P bonding. Interlayer bonding is composed of the covalent bonding between the ($La^{3+}O^{2-}$) layer and the ($Fe^{2+}P^{3-}$) layer. The interaction between the two layers is mainly formed by La and P atoms. This weak La-P hybridization was previously reported by Ref. [15]. Therefore, in the case of the incident beam perpendicular to the c-axis, i.e., q⊥c, such La-P bonding is not the major contribution and excitation from a-b planes is the major contribution. Via careful analysis on the data of DOS and the corresponding transition matrix elements calculated by the Wien 2k software, peak c can be directly interpreted as the transition from La 5s to La 4p states hybridized with P 3p states. This orientation dependence of EELS results basically from the anisotropic feature of the unoccupied electronic structures, because the LaOFeP crystal belongs to a type of two-dimensional layered structure.

The 16.8 eV peak, as shown in Fig. 9(b), could be assigned to the extended O 2s state to the O 2p–La 4p vacant hybridization states. The 28.5 eV peak shows a bump bulk feature. Accordingly, peak b can be identified to be a plasmon peak. To confirm this, collective plasmon excitation has been estimated via the Drude equation of $\hbar\omega_p = \hbar(ne2/\varepsilon_0 m)^{1/2}$ on the basis of free-electron approximation. The density of valence electrons $n$ in LaOFeP is assumed to be about $5.3 \times 10^{26}$ m$^{-3}$, a $\hbar\omega_p$ value of 26.4 eV is obtained. The discrepancy between the theoretical data and the experimental results can possibly be explained by the binding strength of the valence electrons inside LaOFeP crystal. The plasmon peak results from the excitation of all the valence electrons of all the elemental species. Peak d at 55.3 eV is assigned to be Fe $M_{2,3}$ edges, corresponding to the transition from the Fe 3p3/2 and 3p1/2 states to the empty continuum states.

Compared with other techniques, EELS significantly extends the energy range and is a useful technique for analysis of the dielectric properties on a microstructure level. Figure 10 shows the dielectric functions obtained by Kramer-Kronig analysis (KKA) of the spectrum in the low loss energy region. To make the KKA reliable, the contribution from the elastic peak was subtracted and the multiple scattering effects were removed by performing a Fourier-log method. Hence, a single scattering distribution was obtained via this deconvolution process. The imaginary part $\varepsilon_2$ of the dielectric function was obtained by

$$\varepsilon_{2ii} = \frac{4\pi^2 e^2}{m^2(\omega - \Delta_c/\hbar)V} \sum_{v,c,k} |<ck|p_i|vk>|^2 \times \delta(E_{ck} - E_{vk} - \hbar\omega),$$ where $E_{ck}$ and $E_{vk}$ are

the quasiparticle energies approximated to the eigenvalues, $|ck>$ and $|vk>$ are the Bloch functions of the conduction and valence bands, $V$ is the volume of the unit cell, and $p_i$ is the momentum operator with $i=x, y,$ or $z$ direction corresponding to the Cartesian axes. Then, the Kramer-Kronig analysis was performed to calculate the real part $\varepsilon_1$ of dielectric function. The minimum value, maximum value and the intersecting points between $x$ axis and $\varepsilon_1$ curves include the information of interband transition. Peaks in the imaginary part $\varepsilon_1$ and $\varepsilon_2$ are mainly associated with the interband transitions, intraband transitions and collective plasmon excitations.

Figure 11(a) displays the magnetic susceptibility curves for the LaOFeP sample before and after annealing. This annealing was done together with ~0.1 g $La_2O_3$ powder sealed in a silica tube to supply an oxygen source. The strong diamagnetic signal demonstrates the bulk superconducting character in the annealed LaOFeP (sample B). In contrast, no diamagnetic signal could be detected from the LaOFeP product before annealing (sample A), although a decrease of susceptibility value along with decreasing temperature was observed. Figure 11(b) shows Fe $L_{2,3}$ edges core loss EEL spectra for sample A and sample B. Simulated data with energy broadening 0.1 eV are used for comparison. Before comparison, the background of the experimental spectrum was subtracted and multi-scattering effect were deconvoluted. Peaks a and c derive from the spin-orbit splitting Fe $2p_{3/2}$ and $2p_{1/2}$ edges of LaOFeP, respectively. Using the calculation method reported by Wang [25], the ratio of $L_3$ to $L_2$ is estimated to be about 4.4 – 4.6, giving a Fe valence of around 2.0 for LaOFeP, for both annealed and unannealed samples. This Fe valence estimation is consistent with the above calculation.

The most remarkable phenomenon revealed in this spectrum is peak b, to the right side of peak a. Peak b can be observed from the annealed LaOFeP sample and further confirmed by DFT simulations. However, this peak does not show up in the data from the LaOFeP sample without annealing (the top curve), although the X-ray diffraction pattern of both specimens does not show any obvious difference. Detailed simulation analysis on the EELS data and DOS data reveal that peak a corresponds to the transition from Fe 2p to Fe $3d_{z2}$–P 3p hybridized vacant states and peak b corresponds to the transition from Fe 2p to Fe $3d_{x2+y2}$–P 3p hybridized vacant states.

To make the analysis reliable, multiple scattering effects have been deconvoluted from the original spectrum in order to obtain the single scattering spectrum. Accordingly, this single scattering spectrum can be reasonably compared with the electronic DOS data. The basic difference between the LaOFeP sample before and after annealing is that superconductivity can be found at 4.2 K in annealed LaOFeP with but superconductivity cannot be found in LaOFeP without annealing. The occurrence of superconductivity induced by annealing is similar to such phenomena found in High-$T_c$ superconductors, such as $YBa_2Cu_3O_{7-\delta}$ and $La_{2-x}Sr_xCuO_4$ [23-24].

Figure 11(c) shows the oxygen K-edge core loss electron energy loss spectra for LaOFeP before and after annealing. On each spectrum, four evident peaks (a to d) can be observed. Peak a corresponds to the transition from O 1s states towards La $4d_{z2}$–O 2p joint vacant states. The b, c, and d excitations arise chiefly from O 1s states to La $4d_{xy}$–O 2p, La $4d_{xz+yz}$–O 2p, and La $4d_{x2+y2}$–O 2p hybridized states. It should be noted that peak c's contribution becomes more dominant in the superconducting sample with annealing. The effect of annealing on the superconductivity mechanism of LaOFeP is a complicated issue. For example, it is possible that the exact chemical formula of LaOFeP after this annealing might be transformed to LaO$_{1-\delta}$FeP and the carrier density might change. This issue will be addressed in future work. Figure 12 shows the EELS core loss edges of La. White lines can be observed, representing La M$_{4,5}$; and the relative intensity ratio between them is similar to the theoretical calculation data (I$_{L3}$ > I$_{L2}$, which is different from the La$_2$O$_3$ case). Dipole-forbidden M$_2$ and M$_3$ edges also were found at 1100 ~ 1250 eV range, which is not shown here.

The data discussed above reveal the two-dimensional character of LaOFeP revealed by orientation-dependent EELS results. Anisotropic EELS spectra depend on momentum transfer along different orthogonal directions of the selected LaOFeP crystallite sheet. In the *q//c* spectra, the momentum transfer resulting from inter-layer bonding is easily detected. However, in the *q⊥c* spectra, the main momentum transfer is in the La-O layer or the Fe-P layer. Therefore, we observe that the edge at ~55 eV is more intense for momentum transfer q perpendicular to c than parallel to c. By comparison, the layers of Fe-P in the a-b planes of LaOFeP are believed to play a role similar to the CuO$_2$ planes in the high temperature superconducting oxides. From the carrier density point of view, our annealing experiment shows that a low O$_2$ atmosphere is important to the superconductivity of LaOFeP.

## IV. Conclusions:

In summary, we have investigated the electronic band structure and magnetic moment of un-doped and F-doped LaOFeP by first principles calculations. The calculations for NM, FM and AFM states of LaOFeP indicate that the Fe atom does not form a long-range magnetic ordering. The Fe atom in this system shows very weak magnetic moment, which is a necessary condition for the appearance of superconductivity in this material. The Fe magnetic moment is sensitive to the height of the P atom. Analysis of the band structure shows that the charge state of Fe ion in this compound is nearly Fe$^{2+}$. The Fermi surface analysis indicates that there are five FSs corresponding to the Fe 3*d* orbital characteristics, respectively. The virtual crystal approximation method was used to study the effect of F-doping on the electronic structures in LaOFeP. The calculation results indicate that F-doping introduces electrons into Fe $d_z^2$ orbital to adjust the carrier concentration in the FeP layer.

Experimental results of EELS fundamentally agree with the DFT band calculation for the LaOFeP superconductor. We present an interpretation of experimental EELS spectra for the low-loss spectra and core-loss spectra of LaOFeP. Low-energy EELS spectra show that an obvious peak is found at ~44 eV measured with the momentum transfer q//c and not found with the q⊥c spectra. EELS with the direction of momentum transfer implies that this state results from La-P hybridization. Annealing experiments indicate that low oxygen pressure induces the occurrence of superconductivity of LaOFeP, which is evidenced by the O K and Fe $L_{2,3}$ edges. It is supposed that our results might give some reference to the understanding of the superconductivity of LaOFeP.

Acknowledgements
This work is supported by the National Natural Foundation of China and the Ministry of Science and Technology of China (973 Project No. 2006CB301001).

Figure captions:

Figure 1 (Color online)　The calculation model for (a) NM state, (b) FM state and (c) AFM state. The arrows indicate the spin orientation of Fe atoms in each model.

Figure 2 (Color online)　The calculated density of states (DOS) of LaOFeP.

Figure 3(Color online)　(a) Schematic of LaOFeP energy level scheme from -20 eV below $E_F$ to 3.0 eV above $E_F$. (b) The calculated spin up band structure (left) and (c) spin down band structure (right) of LaOFeP. 0 eV denotes the Fermi level. (color online).

Figure 4 (Color online)　The calculated spin up Fermi surfaces of LaOFeP.

Figure 5 (Color online)　The calculated spin down Fermi surfaces of LaOFeP.

Figure 6 (Color online)　The comparison of band structures between the La($O_{0.9}F_{0.1}$)FeP (solid line) and LaOFeP (red dashed line). The top one is the spin up bands near Fermi　level; and the bottom one is the spin down bands near the Fermi level.

Figure 7 (Color online) The calculated spin up Fermi surfaces of LaOFeP.

Figure 8 (Color online) The calculated spin down Fermi surfaces of LaOFeP.

Figure 9 (Color online) (a) Atomic model of LaOFeP along the [1$\bar{1}$0] direction. (b) The low loss EELS spectra with momentum transfer q//c and q⊥c for LaOFeP, respectively. (c) SEM image of a LaOFeP crystal showing its layered structure. (d) HRTEM image of LaOFeP, scale bar=1 nm.

Figure 10 Real part and imaginary part of dielectric function of LaOFeP.

Figure 11 (a) Magnetic susceptibility curves of LaOFeP before and after annealing under low $O_2$ concentration. Electron energy loss spectra of LaOFeP in the energy regions of (b) Fe $L_{2,3}$ ionization edges and (c) oxygen K ionization edges. The calculated data for LaOFeP are shown for comparison. The spectra have been offset in the vertical direction for clarity.

Figure 12 Experimental and simulated electron energy loss spectra of LaOFeP in the region of La $M_{4,5}$ ionization edges.

Fig. 1

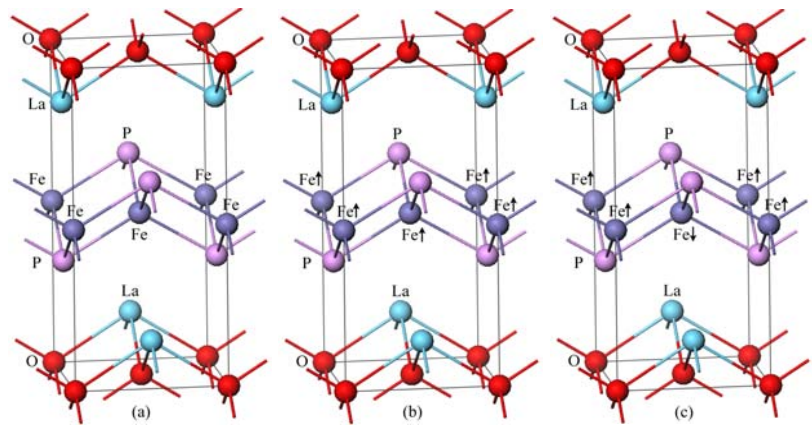

Fig. 2

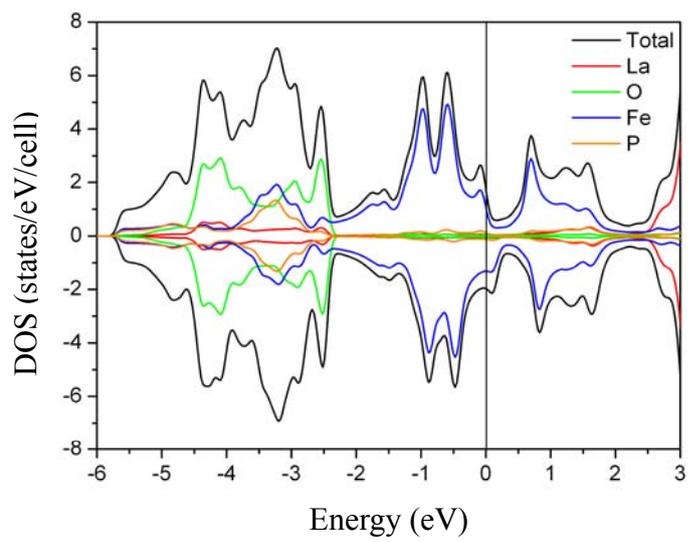

Fig. 3

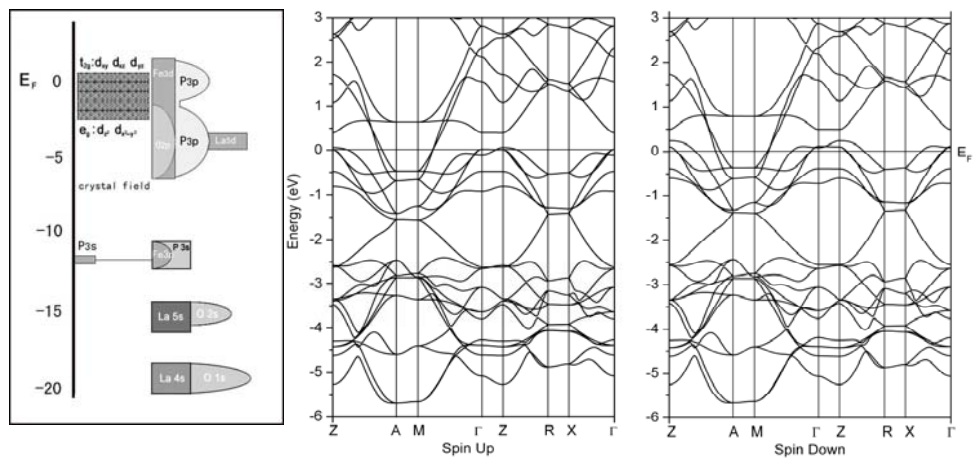

Fig. 4

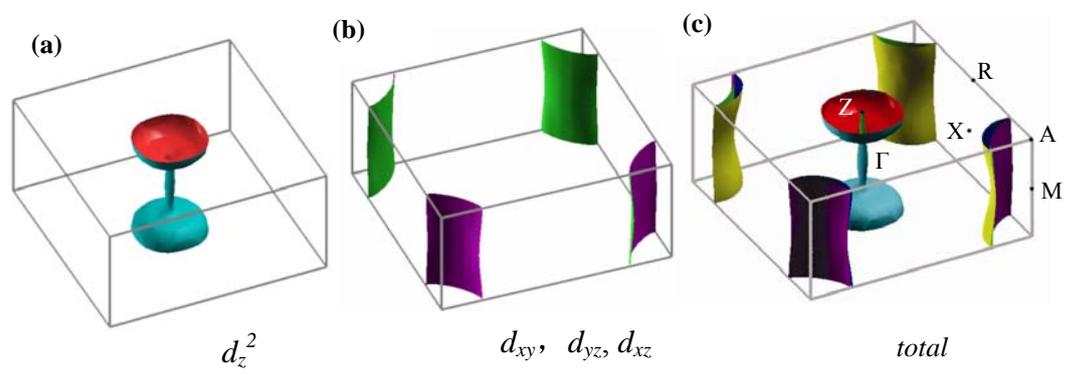

(a) $d_{z^2}$   (b) $d_{xy}$, $d_{yz}$, $d_{xz}$   (c) total

Fig. 5

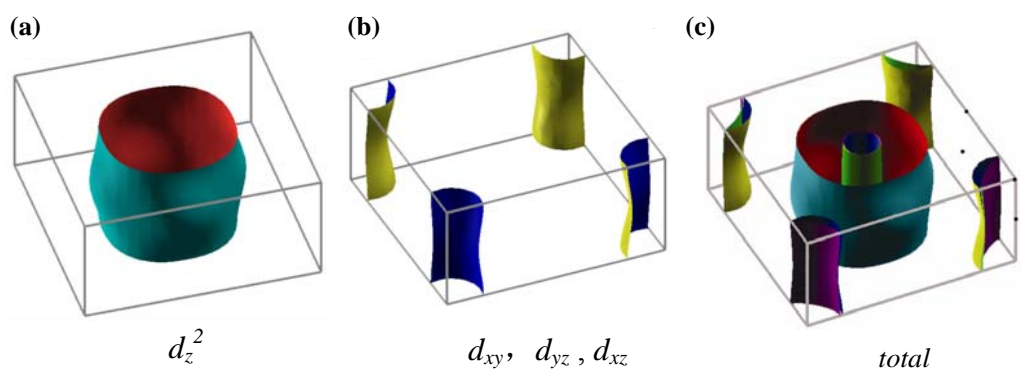

(a) $d_z^2$    (b) $d_{xy}$, $d_{yz}$, $d_{xz}$    (c) total

Fig. 6

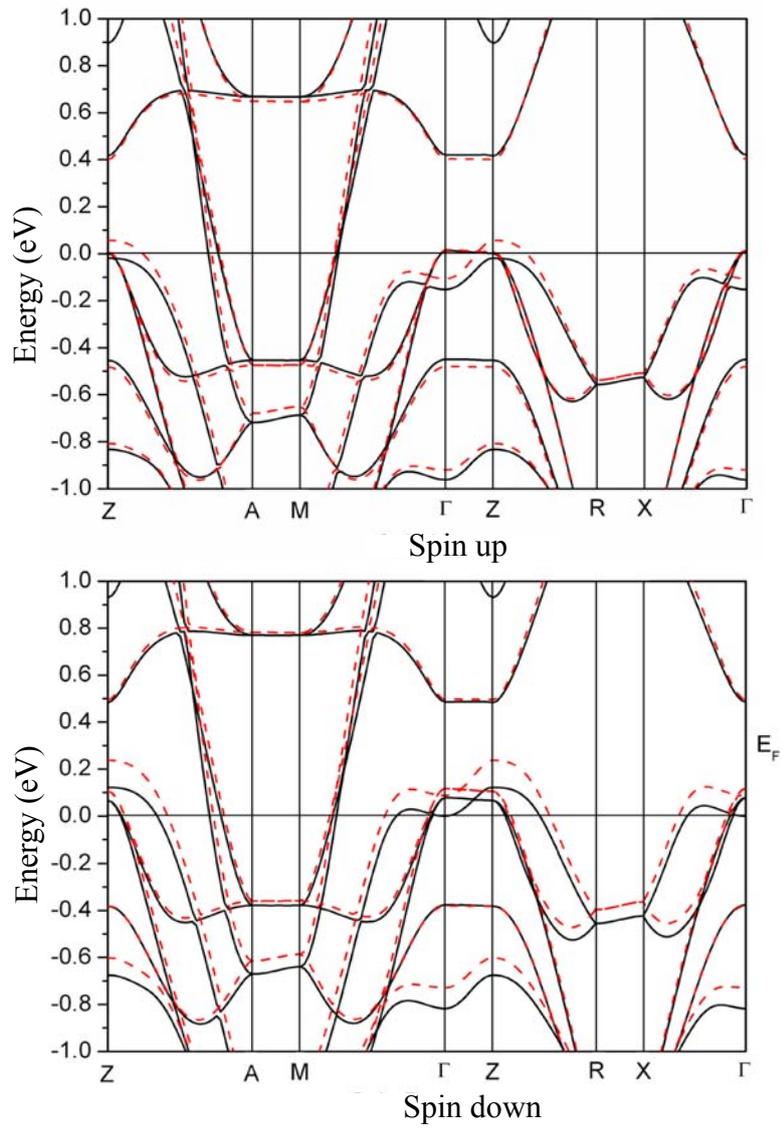

Fig. 7

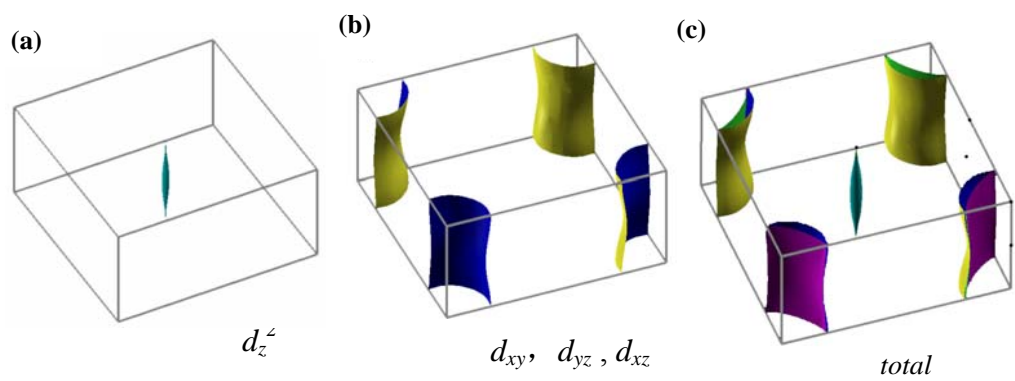

(a) $d_{z^2}$  (b) $d_{xy}$, $d_{yz}$, $d_{xz}$  (c) total

Fig. 8

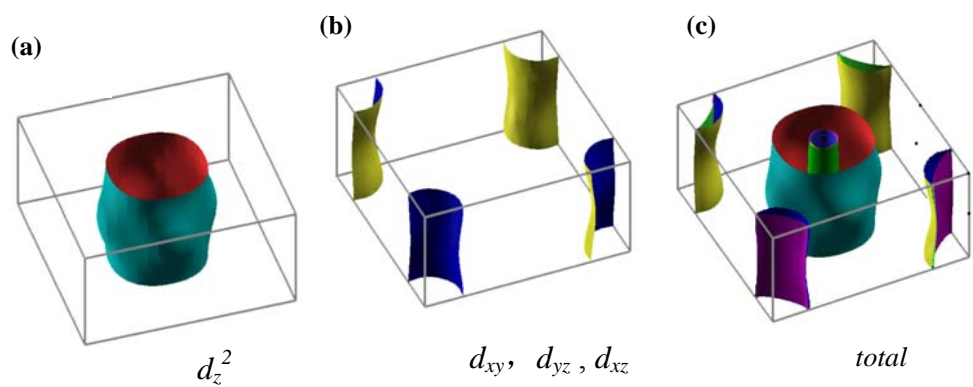

(a) $d_z^2$    (b) $d_{xy}$, $d_{yz}$, $d_{xz}$    (c) total

Fig. 9

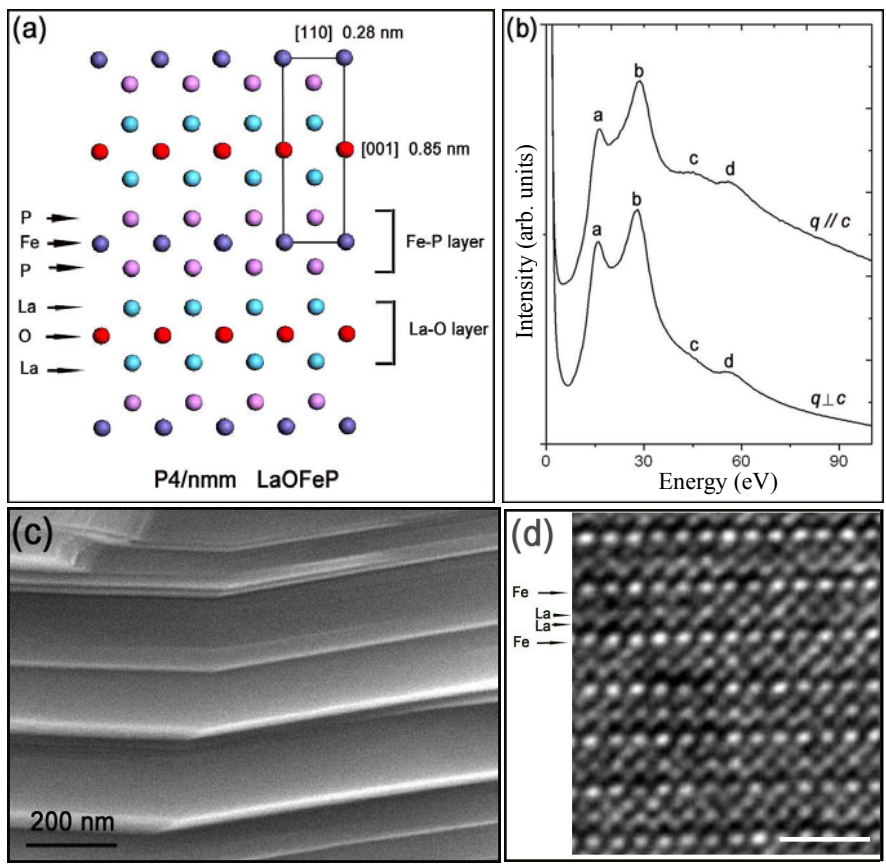

Fig. 10

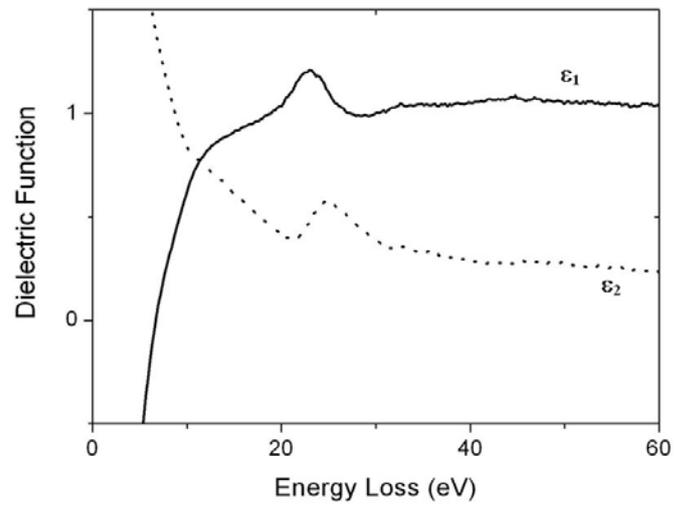

Fig. 11

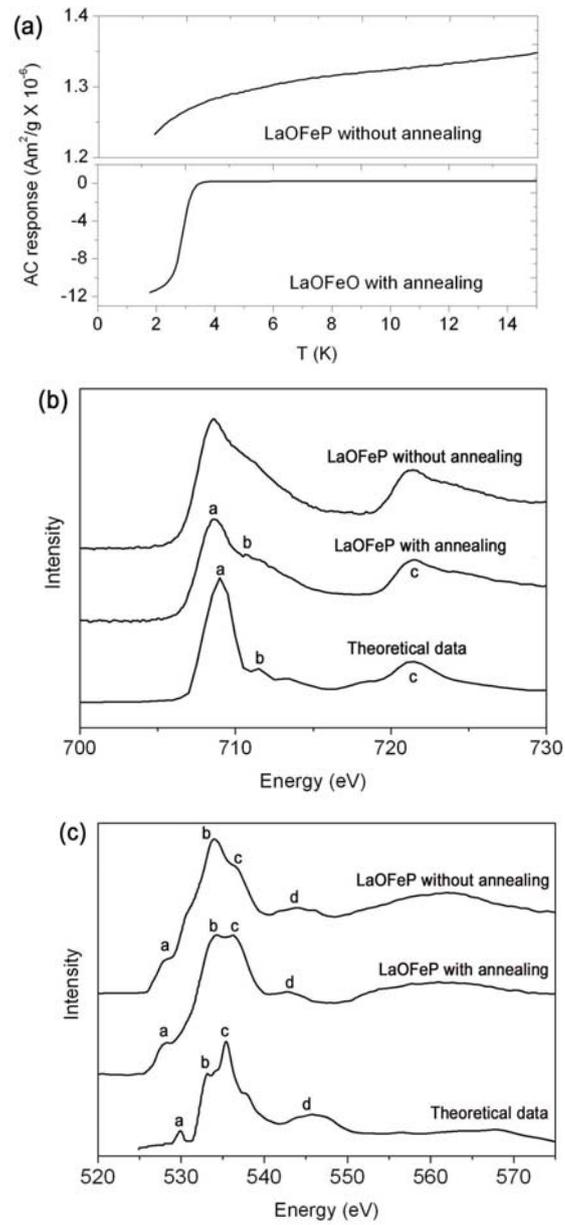

Fig. 12

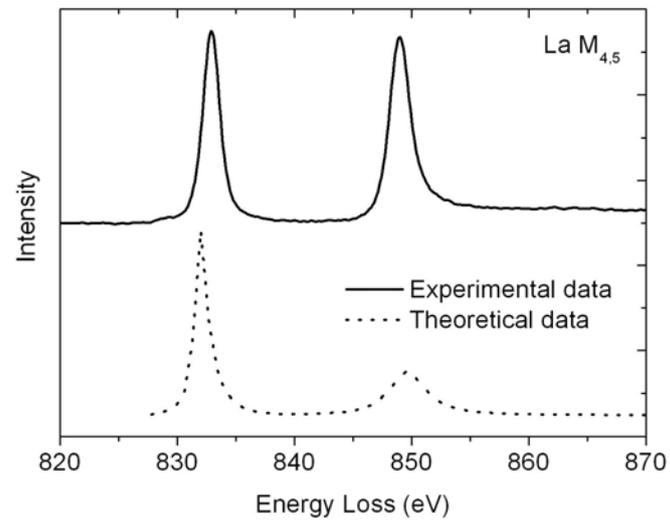